\documentclass[11pt]{article}
\usepackage{graphics}

\begin{document}

\title{\textbf{The World is Either Algorithmic or \medskip Mostly Random}\\Third Prize Winning Essay\\2011 FQXi Contest \emph{Is Reality Digital or Analog?}}
\author{\textbf{Hector Zenil}\\ IHPST\\Universit\'e de Paris 1 -- Panth\'eon-Sorbonne
}

\date{}

\maketitle

\begin{abstract}
\noindent I will propose the notion that the universe is digital, not as a claim about what the universe is made of but rather about the way it unfolds. Central to the argument will be the concepts of symmetry breaking and algorithmic probability, which will be used as tools to compare the way patterns are distributed in our world to the way patterns are distributed in a simulated digital one. These concepts will provide a framework for a discussion of the informational nature of reality. I will argue that if the universe were analog, then the world would likely be random, making it largely incomprehensible. The digital model has, however, an inherent beauty in its imposition of an upper limit and in the convergence in computational power to a maximal level of sophistication. Even if deterministic, that it is digital doesn't mean that the world is trivial or predictable, but rather that it is built up from operations that at the lowest scale are very simple but that at a higher scale look complex and even random, though only in appearance.
\end{abstract}


\section{Everything out of nothing}

Among the simplest hypotheses compatible with the best account of the origin of the universe that we currently have, one can either choose to start from nothing, the state of the universe with all its matter and energy squeezed into an infinitely small point of no length, no width, and infinite density called a singularity; or else a fraction later, out of a state of complete disorder such that once particles formed they couldn't do anything except collide with each other in a completely disordered way. In either case there had to have been a transition to the state in which we find ourselves today, in a universe with physical laws describing our reality, from the biggest to the tiniest, laws that are often simple enough to be easily comprehensible even if the phenomena they describe are complicated. The universe seems highly ordered and structured today, in contrast to the background noise left behind by the Big Bang, which is similar to what one would see on the screen of an old untuned analog TV (Fig.  1, left image) or hear on an untuned radio station. Given the random state physicists believe to have existed at the inception, embodying little to no information, how did we arrive at the structured world that we now inhabit?

\begin{figure}[htdp]
\centering
   \scalebox{.3}{\includegraphics{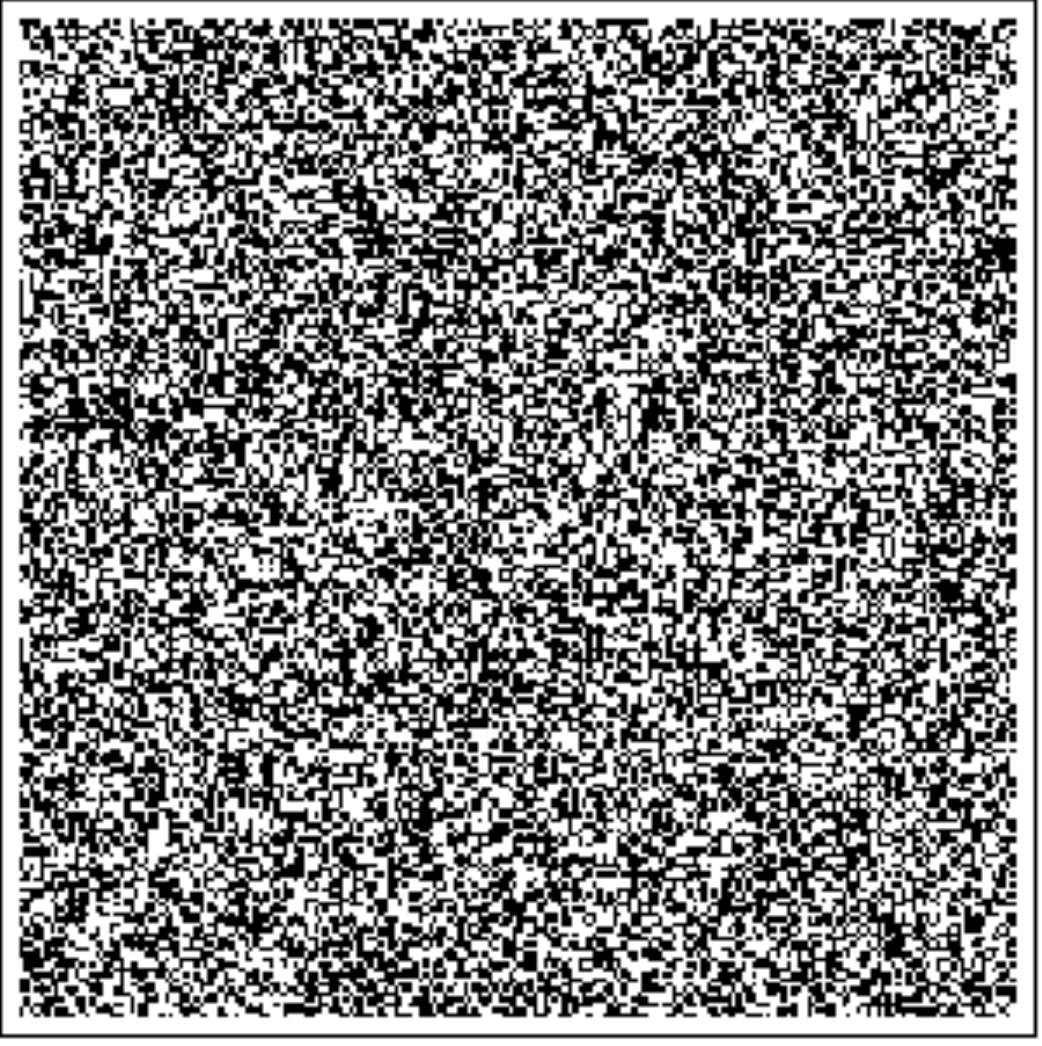}}\scalebox{.37}{\includegraphics{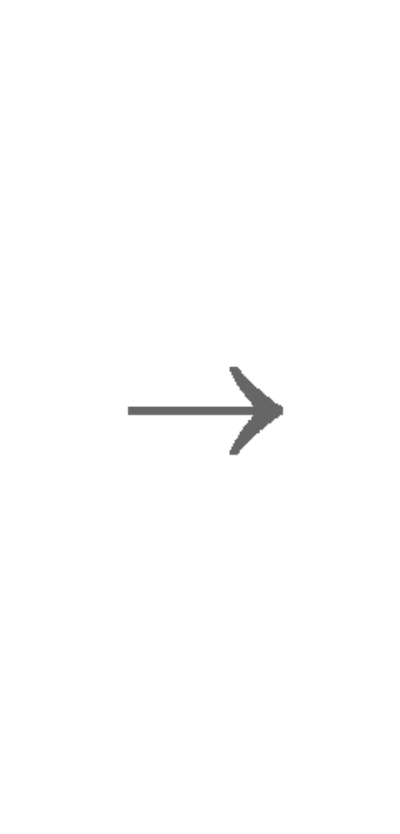}}\scalebox{.38}{\includegraphics{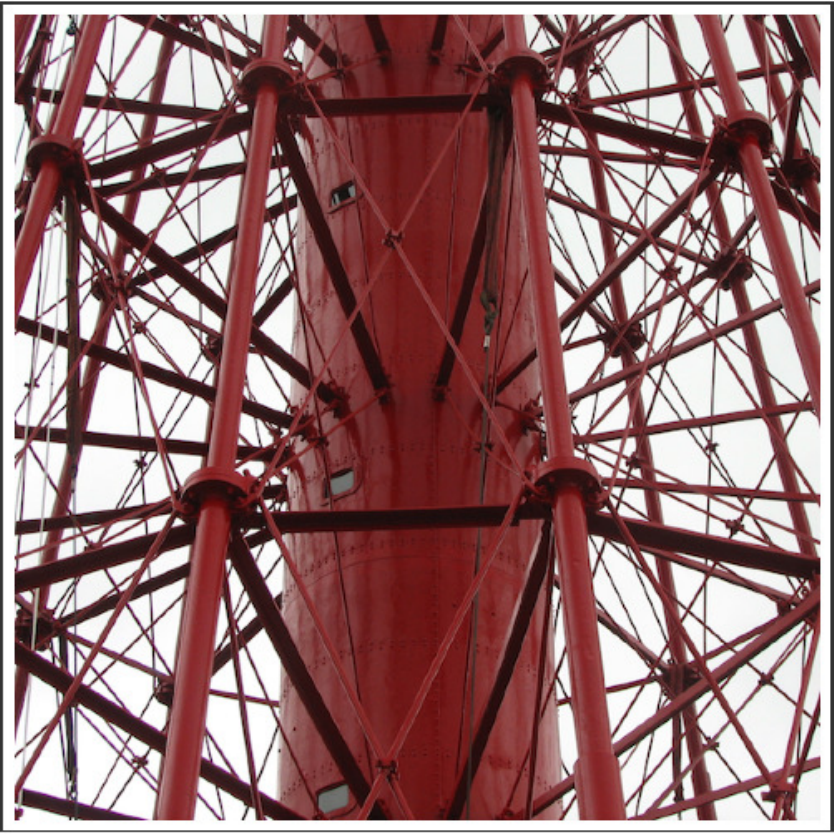}}
\caption{\footnotesize{From noise to highly organized structures. The cosmic background radiation (left) or what the universe looked like at all scales in every direction, and the kinds of structures (right) we find in our everyday life today.}}
\end{figure}

Whether the universe began its existence as a single point, or whether its inception was a state of complete randomness, one can think of either the point or the state of randomness as  quintessential states of perfect symmetry. Either no part had more or less information because there were no parts or all parts carried no information, like white noise on the screen of an untuned TV. In such a state one would be unable to send a signal, simply because it would be destroyed immediately. But thermal equilibrium in an expanding space was unstable, so asymmetries started to arise and some regions now appeared cooler than others. The universe quickly expanded and began to produce the first structures.

When the universe had cooled to the point where the simplest atoms could form, white noise no longer dominated and matter took over, embarking on a process of structure formation. The first symmetry breaking led to more symmetry breaking. This symmetry breaking can be found everywhere in the universe, from the great disparity between matter and antimatter (atoms with inverted charge particles) to the way planets rotate in a single direction around the sun; or in the form of what is today known as homochirality, groups of molecules that lack a configuration of non-superposable mirror images; and in living beings, all of whom share amino acids and sugars but genetically encoded so that each possesses one particular (arbitrary) molecular orientation rather than another. This symmetry breaking is the fabric of information. The laws of physics may have arisen in this way, not as agents shaping the universe but as a result of the unfolding of this dynamic of information processing from symmetry breaking.

\subsection{Complexity from randomness}

If you wished to produce the digits of the mathematical constant $\pi$ by throwing digits at random, you'd have to try again and again until you got a few consecutive numbers matching an initial segment of the decimal expansion of $\pi$. The probability of succeeding would be very small:  $1/10$ multiplied by the desired number of digits. For example, $(1/10)^{2400}$ for a segment of length 2400 digits of $\pi$. But if instead of throwing digits into the air, one were to throw bits of computer programs and execute them on a digital computer, things turn out to be very different. For example, a program producing the digits of the mathematical constant $\pi$ would have a greater chance of being produced by a computer program. The following is an example of a program written in ANSI C language of only 158 characters producing the first 2400 digits of $\pi$:

\begin{verbatim}
int a=10000,b,c=8400,d,e,f[8401],g;main(){for(;b-c;)
f[b++]=a/5;for(;d=0,g=c*2;c-=14,printf(``\%.4d'',e+d/a),
e=d\%a)for(b=c;d+=f[b]*a,f[b]=d\%--g,d/=g--,--b;d*=b);}
\end{verbatim}

This program compresses the first 2400 digits of $\pi$, and there are many other formulae that can be implemented as short computer programs to generate any arbitrary number of digits of $\pi$. 

Computer programs are like physical laws; they produce order by filtering out a portion of what one feeds them. Start with a random-looking string and run a randomly chosen program on it, and there's a good chance your random-looking string will be turned into a regular, often non-trivial, and highly organized one. In contrast, if you were to throw particles, the chances that they'd group in the way they do if there were no physical laws would be so small that nothing would happen in our universe. Physical laws, like computer programs, make things happen.

Just as formulae producing the digits of $\pi$ are compressed versions of $\pi$, physical laws distill natural phenomena from a series of observations. These laws are valuable because thanks to them one can predict the outcome of a natural phenomenon without having to wait for it to unfold in real time. Solve the equations describing planetary motion and instead of having to wait two years to know the future positions of a planet, one can (almost\footnote{Although this is a whole subject unto itself, it may be pointed out here that the fact that our theories are approximate is due to the same symmetry-breaking happening at all scales, making our predictions diverge in the long term. But it is this same phenomenon---a phenomenon that one may associate with imperfection as opposed to perfect symmetry--that has continuously created information in the past and continues to do so still.}) know them precisely and in a fraction of a second (the time it takes to compute the equation) two years in advance. Laws are always associated with calculations, and it is no coincidence that all these calculations turn out to be computable, whether the computations are carried out by humans or by computers. For all practical purposes physical laws are just like computer programs.

\section{A bit-string Universe}

When Leibniz started doing binary arithmetic, he thought that the world could have come into existence from nothing (zero) and one because everything could be written in this language (the simplest possible language by number of symbols involved), the binary language. Out of this extremely simplified language, out of its beauty, the world came into being before his eyes. If the world were digital at the lowest scale one would end up seeing nothing but strings of bits. They would have looked very random in the first seconds of existence of the universe, and they would suddenly have started displaying patterns equivalent to clusters of particles--as actually happened. Strings would have formed the equivalent of galaxies, bit-string galaxies, bit-string solar systems and bit-string planets, indeed all manner of bit-string things.

The existence of human-made digital computers in the universe is an obvious demonstration that the universe is capable of performing digital computation. The main questions therefore are whether this computation occurs naturally in the universe, how pervasive it is, and whether it is of the same kind as that performed by digital computers. In other words, how different is a bit-string universe to the universe in which we live. The bit-string version is of course an oversimplification, but the two may not, in the end, be that different. 

For if all matter is made of the same basic particles, what makes one object different from another other than the fact that it occupies a different space? What makes a cup a cup and not a human being is quite simply the way its particles are configured. One could disassemble a cup (or several cups) and reassemble them as a human being. What makes a cup a cup and a human being a human being is information. In an informational universe, the world would be computing itself, enabling things to remain themselves. Our digital computers would be reprogramming a part of the universe to make it compute what we want it to compute.

\begin{figure}[htdp]
\centering
   \scalebox{.23}{\includegraphics{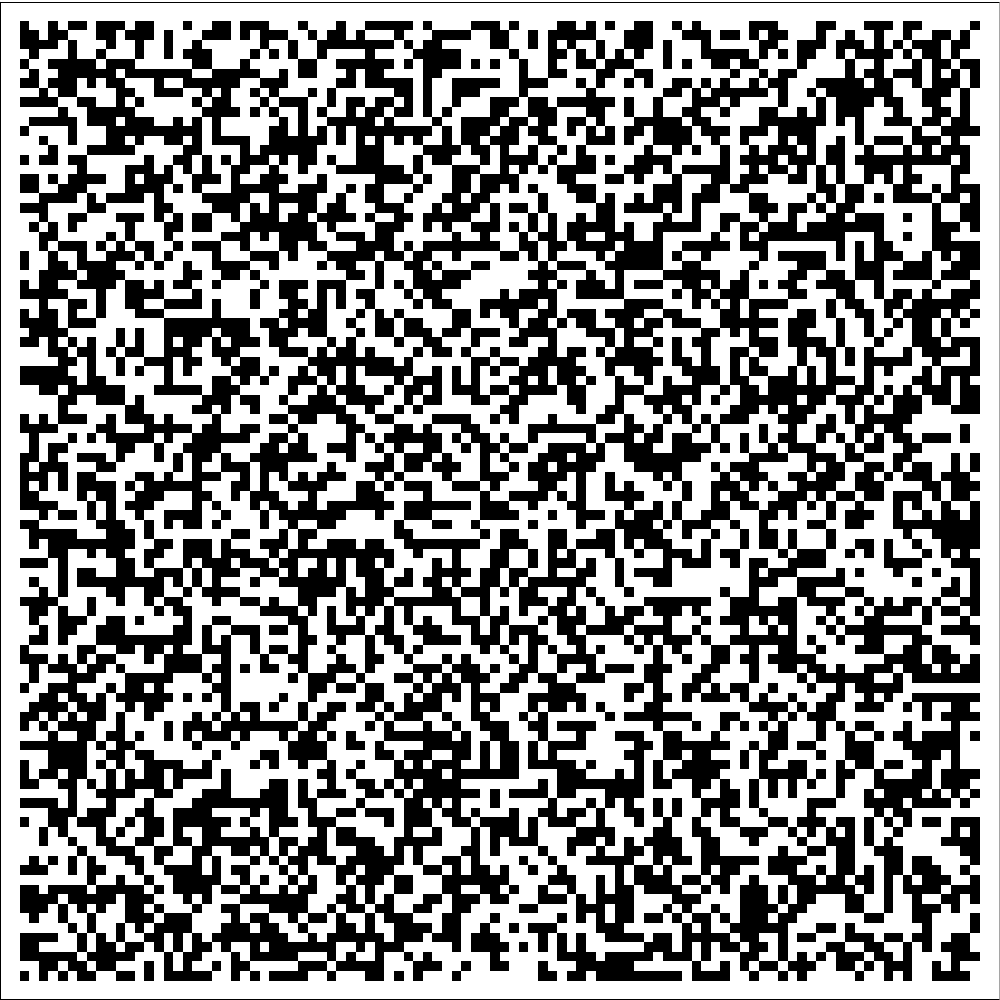}}\scalebox{.2}{\includegraphics{arrow-eps-converted-to.pdf}}\scalebox{.23}{\includegraphics{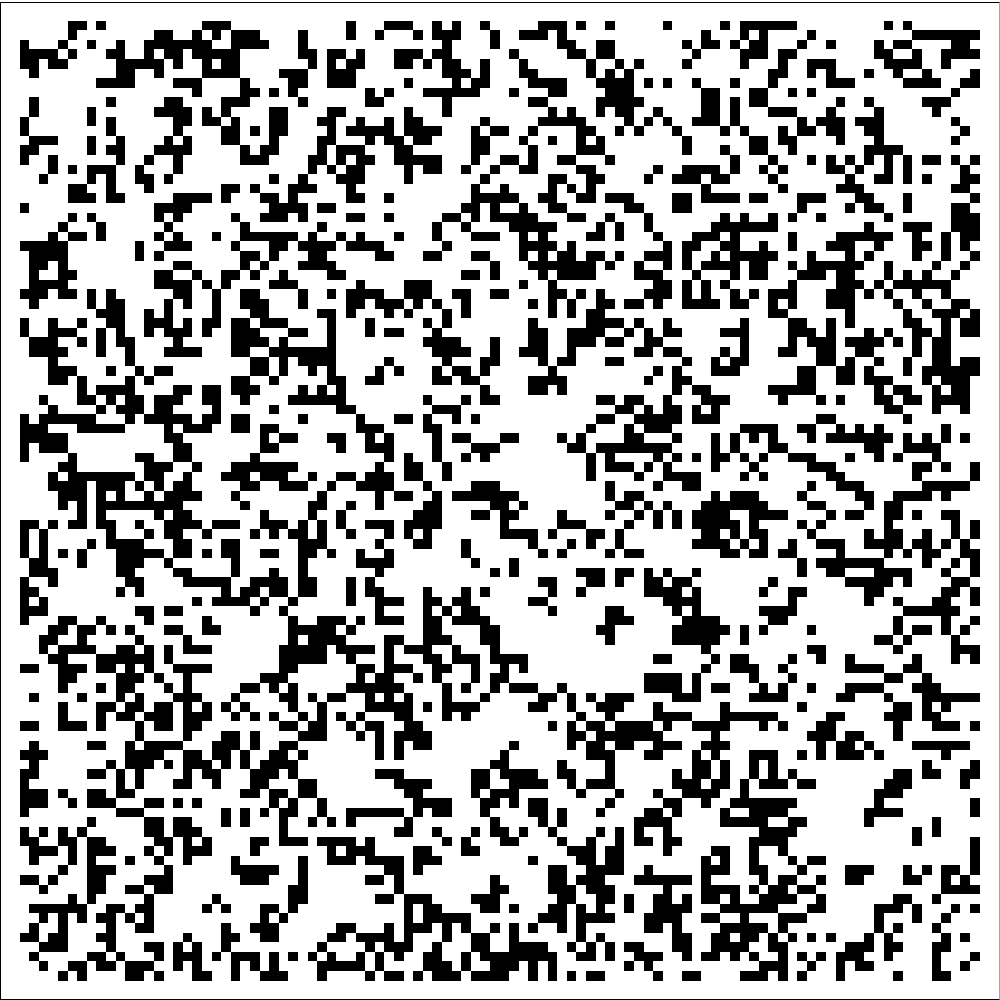}}\scalebox{.2}{\includegraphics{arrow-eps-converted-to.pdf}}\scalebox{.23}{\includegraphics{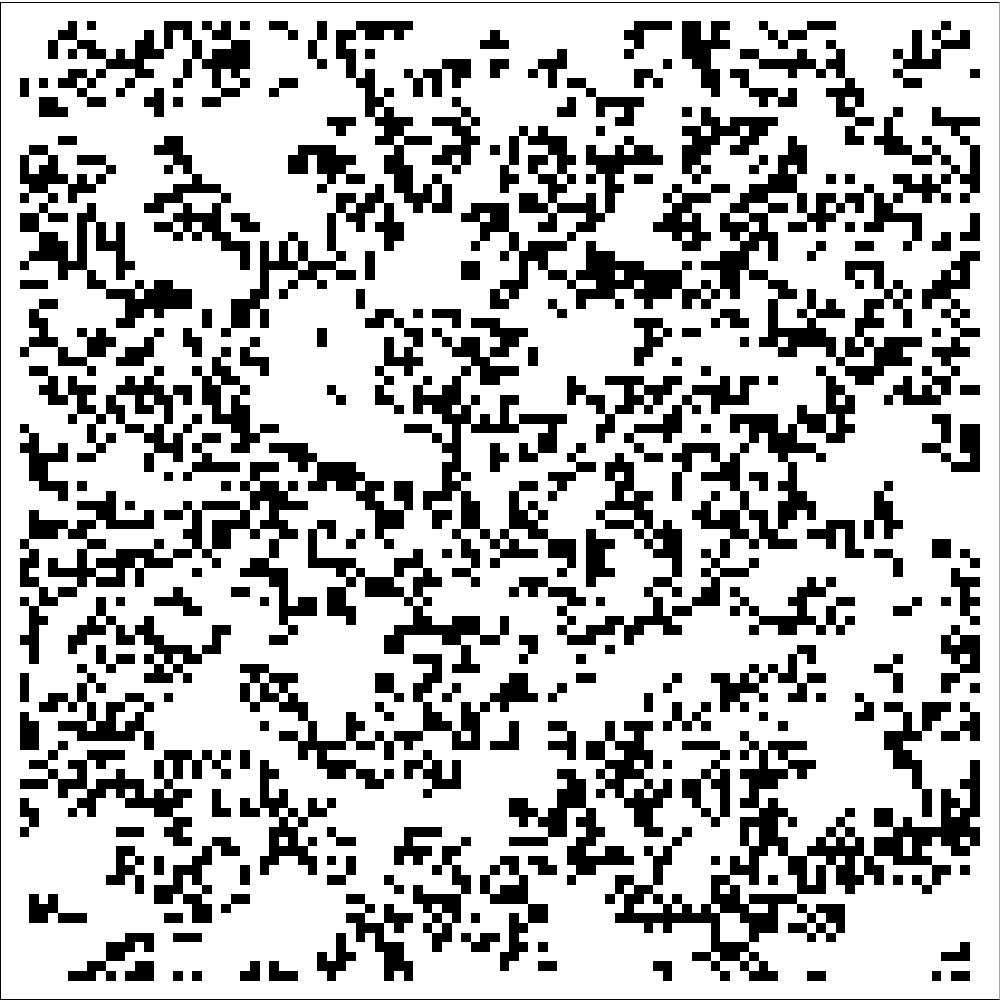}}\scalebox{.2}{\includegraphics{arrow-eps-converted-to.pdf}}\scalebox{.23}{\includegraphics{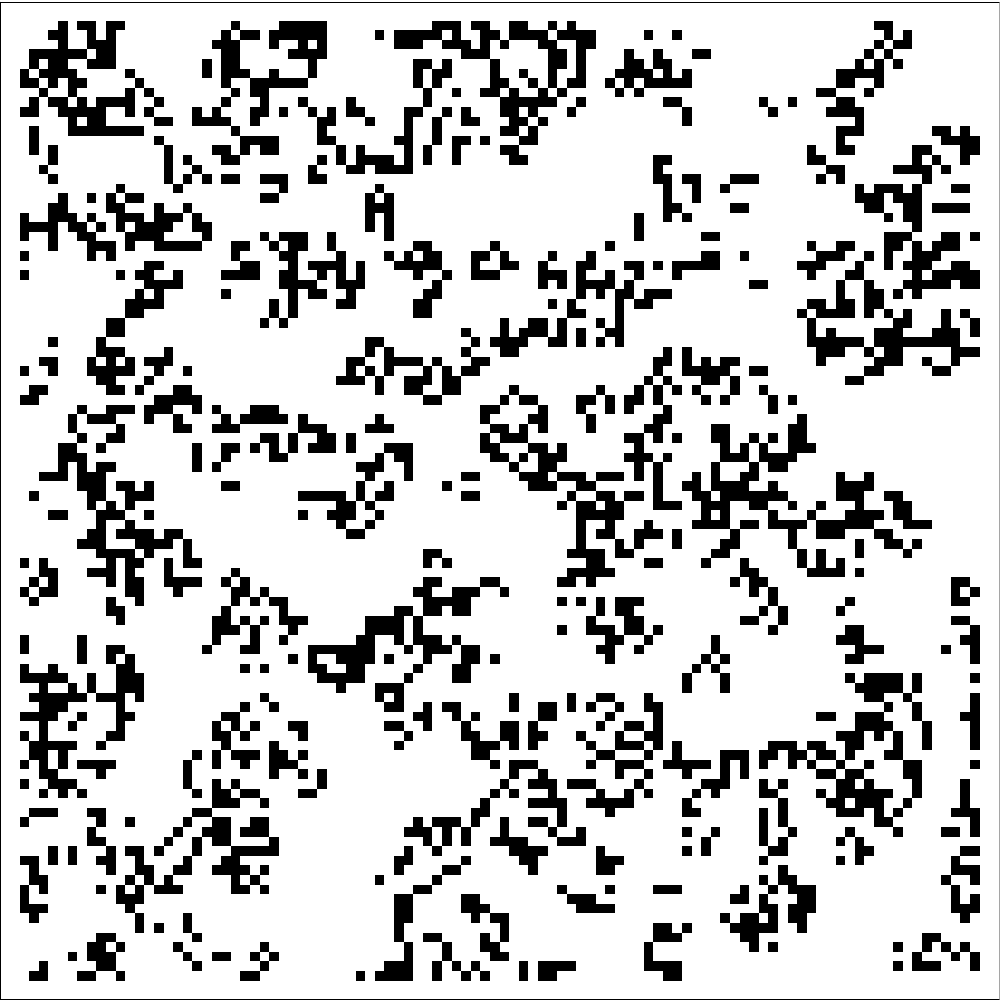}}
\caption{\footnotesize{Patterns generated by a 2D 9-neighbor with Wolfram's rule number 40. Shown here are state space diagrams every 6 runtime steps starting from a $100\times100$ array of (pseudo)random bits.}}
\end{figure}

But at a microscopic level, the quantum greatly differs from the macroscopic world. In the macroscopic world randomness is apparent, according to this view (and ultimately according to classical mechanics), but it is fundamentally different under the standard interpretation of quantum mechanics. At the quantum scale, things seem disconnected, yet a compatible informational interpretation may be possible. Information can only exist in our world if it is carried by a process; every bit has to have a correspondent physical carrier. Even though this carrier is not matter, it takes the form of an interaction between components of matter, an atom interacting with another atom, or a particle interacting with another particle. But at the lowest level, the most elementary particles, just like single bits, carry no information  (the Shannon Entropy of a single bit is 0) when they are not interacting with other particles. They have no causal history because are memoryless isolated from external interaction. When particles interact with other particles they link themselves to a causal network and look as if they were forced to define a value as a result of this interaction (e.g. a measurement). What surprises us about the quantum world is precisely its lack of the causality that we see everywhere else and are so used to. But it is the interaction and its causal history that carries all the memory of the system, with the new bit appearing to us as if it had been defined at random. Linking a bit to  the causal network may also produce correlations of measurements between seemingly disconnected parts of space, but if space is informational at its deepest level, if information is even more fundamental than the matter of which it is made and the physical laws governing that matter, the question of whether these effects violate physical laws may be irrelevant. 

As Charles Bennett\cite{bennett} has pointed out before, in a discrete universe, information is never lost. If one throws a deck of cards into the air, they spread out as they fall, the original order becomes unknowable once on the floor, but the information indicating the cards exact original position and path to the floor is hidden in the form of the air molecules that the cards displaced, the particles that the cards touched in their way to the floor, the particles touched by the particles touched by the cards, even the heat all this particles friction produced. Reversing every motion would restore the deck to its initial condition. If anything in the chain turns out to be random this basic principle collapses, how much the collapsing would affect the world depends of how the world is connected at all scales. Whether there is a kind of garbage collector at the lowest level of the universe to erase this memory is yet to be discovered but as Bennett points out based in the principles of thermodynamics, this is not the case, information does not disappears. If the world were analog the way this information dissipates may allow information loss. But producing random bits in a discrete universe, where all events are the cause of other events, would actually be very expensive because one would need to devise a way to break the causal network, assuming that this were possible to begin with, produce a random bit, and keep the rest of the causal network untouched (otherwise we would see nothing but randomness, which is not the case).

\section{The algorithmic nature of the world}

If one does not have any reason to choose a specific distribution and no prior information is available about a dataset, the uniform distribution is the one making no assumptions according to the principle of indifference. Consider an unknown operation generating a binary string of length $k$ bits. If the method is uniformly random, the probability of finding a particular string $s$ is exactly $2^{-k}$, the same as for any other string of length $k$, which is equivalent to the chances of picking the right digits of $\pi$. However, data (just like $\pi$--largely present, for example, in nature in the form of common processes relating to curves) are usually produced not at random but by a process. 

There is a measure which describes the expected pattern frequency distribution of an abstract machine running a random program. A process that produces a string $s$ with a program $p$ when executed on a universal Turing machine $T$ has probability $m(s)$\cite{levin} (identified as the miraculous universal distribution in \cite{kirchherr}). For any given string $s$, there is an infinite number of programs that can produce $s$, but $m(s)$ is defined such that one can assign a probability of a string being produced by a random program (see the Appendix). 

The distribution $m(s)$ has another interesting particularity, one can start out of almost anything and, as most probabilistic distributions, the distribution remains mostly unchanged. It is the process that determines the shape of the distribution and not the initial conditions from which the programs may start from. This is important because one does not make any assumption on the distribution of initial conditions but on the distribution of programs. Programs running on a universal Turing machine should be uniform, which does not necessarily mean truly random. For example, to approach $m(s)$ from below, one can actually define a set of programs of certain size and define any enumeration to systematically run each program one by one.

This is important because this means one does not actually need \emph{true} randomness, the kind of randomness assumed in quantum mechanics. So one does not really need quantum mechanics to explain the complexity of the world or to underly reality to explain it, one does require, however, computation, at least in this informational worldview. It is information that we think may explain some quantum phenomena and not quantum mechanics what explains computation (neither the structures in the world and how it seems to algorithmically unfold), so we put computation at the lowest level underlying physical reality.

Just as strings can be produced by programs, we may ask after the probability of a certain outcome from a certain natural phenomenon, if the phenomenon, just like a computing machine, is a process rather than a random event. If no other information about the phenomenon is assumed, one can see whether $m(s)$ says anything about a distribution of possible outcomes in the real world. In a world of computable processes, $m(s)$ would indicate the probability that a natural phenomenon produces a particular outcome and how often a certain pattern would occur. If you were going to bet against certain events in the real world, without having any other information, $m(s)$ would be a reasonable decision if the world were digital, just as $m(s)$ is for abstract (digital) machines.

\begin{figure}[htdp]
\centering
   \scalebox{.4}{\includegraphics{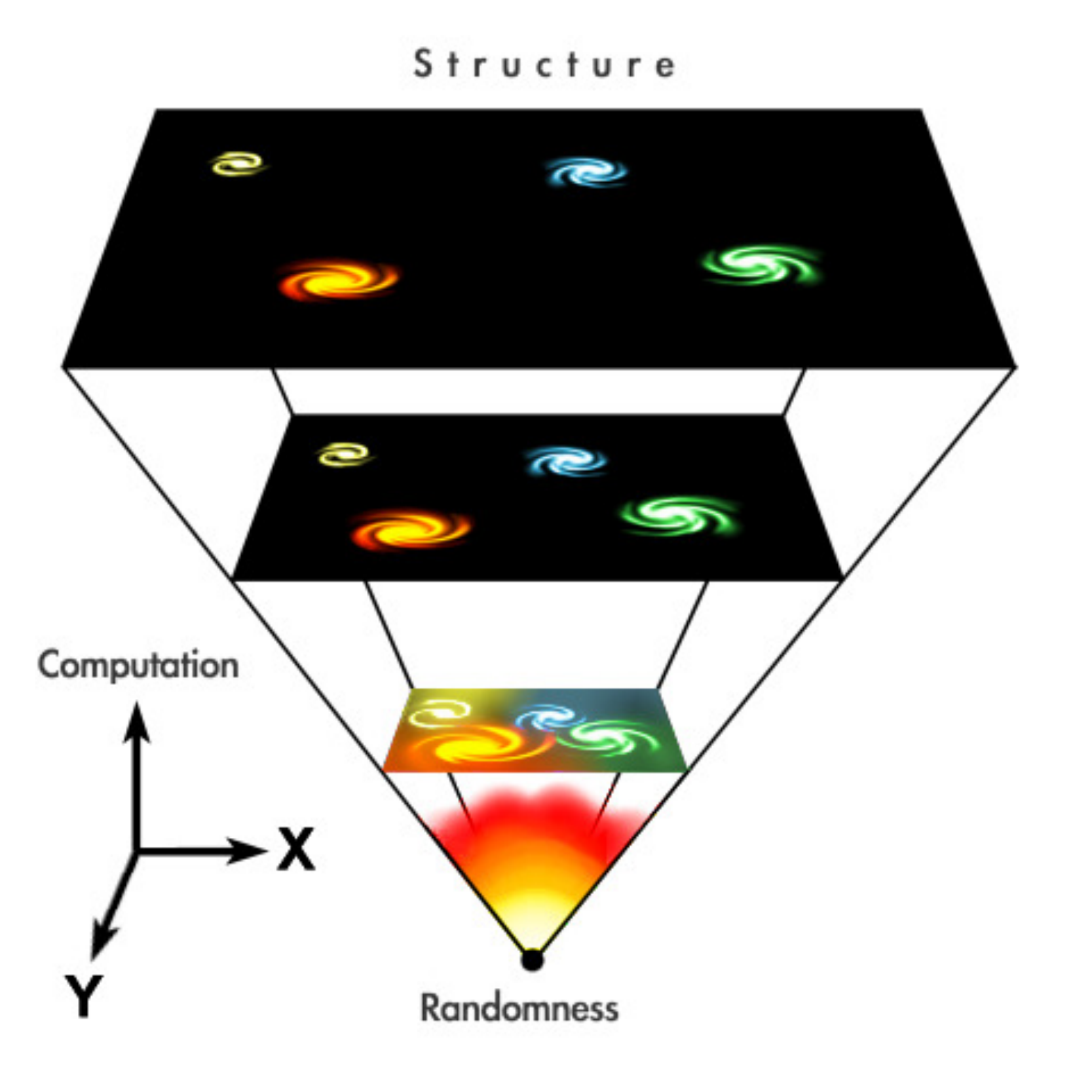}}
\caption{\footnotesize{Start out of nothing or out of randomness and one gets a structured world by iterated computation.}}
\end{figure}

\subsection{Unveiling the machinery}

So if one wished to know whether the world were algorithmic in nature, one would first need to specify what an algorithmic world would look like. If the world is in any respect an algorithmic world, the structures in it should be alike, with their distribution of patterns resembling each other. To demonstrate this, we conceived and performed\cite{zenil} a series of experiments to produce data by purely algorithmic means in order to compare it to sets of data produced by several physical sources. 

On the one hand, samples from physical sources were taken. At the right level data can always be written in binary because each physical observation leading to values (weight, location, etc.) can be numerated independently as a discrete sequence. Each frequency distribution is the result of the count of the number of occurrences of $k$-tuples (substrings of length $k$) extracted from the binary strings from both the empirical and the digital datasets obtained from the output of digital machines. On the other hand, we produced an experimental version of $m(s)$ by running a large set of small Turing machines for which the halting times are known (thanks to the Busy Beaver problem).  The frequency distributions generated were then statistically compared (see \cite{zenil}). A ranking correlation test was carried out and its significance measured to validate either the null hypothesis or the alternative (the latter being that the similarities among empirical data and the digital world are due to chance). 

Some similarities and differences between the physical world and the purely algorithmic world were found. To cite a typical example of a difference, in  the study of the string symmetry group (see Appendix), strings of the type $(01)^n$ appear low-ranked in the empirical data distributions, unlike in the digital ones, where they appear better ranked. In the real world, just as processes don't start from a blank tape (to use a Turing machine as an analogy--see Appendix), strings are usually not delimited, one cannot really tell when a physical phenomena has stopped but rather when the measurement finished. In the real world, the probability of destroying highly symmetrical strings due to this lack of delimitation is higher than that of assembling a symmetrical string by changing bits at random. There is no way to tell when a process starts or ends in nature, and likewise there is no way to set measurements taking into account the ``right'' time lengths of empirical dataset streams.

In the simulated world things are different because machines have a halting configuration, such as a special halting state. And because in our simulated world machines have no interaction with any other machines, periodic and highly symmetric strings have a greater chance of making it intact, appearing better ranked in the frequency of strings from higher to lower frequency (hence lower to higher random complexity). To verify that this was actually the case we set up two other different experiments. One consisted in starting the machines from random initial conditions. Even if this does not fully simulate having processes interacting with other processes at every step, it is a setup closer to what happens in the real world, where computations usually start where other computations end. The other experiment consisted of running non-self-delimited machines like one-dimensional cellular automata which by definition do not have any halting configuration. Their computations could be halted at arbitrary times, resulting in strings of arbitrary length, just as would happen in most experiments in the real world when one has basically to decide when to stop making a measurement in order to start a frequency analysis. What we found was what we were expecting to find: the highly structured $(01)^n$ string was ranked lower compared to the first simulation, and the longer the $n$ the less well ranked in comparison to the original experiment with halting machines, and closer to the rank of the distributions of empirical data. 

What happens is that in the real world highly organized strings have little chance of making it if they interact with other systems. Changing a single bit destroys a perfect 2-period pattern of a $(01)^n$ string, and the longer the string, the greater the odds of it being destroyed by an interaction with another system or because it has been trimmed in the wrong place at the time of the measurement. In the case of halting Turing machines, however, strings are delimited by the code telling the machine to produce exactly $n$ alternations and halt.

We found that while the correlation between the real world dataset and the digital dataset was not significant enough to be conclusive, each distribution was correlated--with varying degrees of confidence--with at least one algorithmic distribution produced by a model of computation. In other words, distributions from empirical data disperse patterns in a similar way to the distributions generated by using digital computers (Turing machines, cellular automata and other abstract machines).

\subsection{How different is our world from a simulated digital one?}

To illustrate how natural processes may be algorithmic, and to specify what we really mean by this, the case of DNA could serve as a perfect example. Processes known to be involved in the replication and transmission of DNA are, among others, chromosomal translocation (a fragment of one chromosome is broken off and is then attached to another), reverse transcriptase, fragment code exchange and crossover, chemical annealing (pairing by hydrogen bonds to a complementary sequence) and DNA denaturation (separation into single-stranded lengths through the breaking of hydrogen bonding between the bases). These are all relatively simple processes. A subset of purely digital operations can match such operations with computational ones. Operations such as joining, copying, partitioning, complementation, trimming, or replacing are equivalent to those that may be observed in DNA. Rules determining the way DNA replicates may also be of the same algorithmic type as those governing other types of physical phenomena, leading us to sometimes discover strong similarities in their tuples distribution. DNA construction, except perhaps for the mutation operation (the result of interaction with another system and not necessarily a true random operation) is the result of a long period of application of simple rules, with layer upon layer of the code of life built up over billions of years in a deep algorithmic process with its own characteristic rules making things like protein unfolding to appear to us highly complex.

The claim that empirical data can be treated as a whole as a distribution (i.e. the frequency of certain patterns against others)--as if all empirical data were of the same nature--must of course be made with great circumspection. One would first need to show that there is a general joint distribution behind all sorts of empirical data. This was indeed something that we tested and reported\cite{zenil}. The demonstration that most empirical data carries an algorithmic signal is that most data are comprehensible to some greater or lesser degree. Think of the diverse kinds of data that you store in your personal computer, whether music, images or text, all of them highly (lossless) compressible.

One may wonder whether the lossless compressibility of data is in any sense an indication of the discreteness of the world. It is, and we have presented some material here that supports our answer. The relationship is actually strong; the chances of finding incompressible data in an analog world are much greater, simply because the possibilities for anything are much greater. The essential difference between one world and the other is the introduction of actual infinity. An analog world means that one can divide space and/or time into an infinite number of pieces, and that matter and everything else may be capable of following any of these infinitely many paths and convoluted trajectories. It may seem possible to make a lot of sense out of it by way of the successful fields of differential calculus and mathematical analysis, yet one has to make a distinction between what symbols represent and what the actual calculations among the symbols are.

If our world is analog, it would have a greater chance of looking like a Chaitin $\Omega$, a random number by definition (see the Appendix), dependent on the unpredictability of universal digital computers, also called the halting probability. The world may be a small, apparently ordered fragment of a globally random universe, as Calude and Meyerstein have suggested\cite{caludemeyer}. In which case we should consider ourselves extremely lucky to live in a tiny, apparently ordered part.

Unlike some physicists who seem to think that a theory explaining the universe will ultimately be very complicated and mathematical (for example the views expressed by Stephen Weinberg in a recent interview with Amir Aczel\cite{pls}), we think that the correlations found are due to the following reason: general physical processes are dominated by simple algorithmic rules, the same rules that digital computers are capable of carrying out. Our approach suggests that the information in the world is the result of processes resembling computer programs rather than of dynamics characteristic of a more random, or analog, world.

\newpage

\section*{Appendix: Some basic theory}

\subsection*{Turing machine}

A Turing machine\cite{turing} is a $5$-tuple $\{s_i,k_i,s_i^\prime,k_i^\prime,d_i\}$, where $s_i$ is the tape symbol the machine's head is scanning at time $t$,  $k_i$ the machine's current state (the instruction) at time $t$, $s_i^\prime$ a unique symbol to write (the machine can overwrite a 1 on a 0, a 0 on a $1$, a $1$ on a $1$, or a $0$ on a $0$) at time $t+1$, $k_i^\prime$ a state to transition into (which may be the same as the one it was already in) at time $t+1$, and $d_i$ a direction to move in time $t+1$, either to the right ($R$) cell or to the left ($L$) cell, after writing. At a time $t$ the Turing machine produces an output described by the contiguous cells in the tape visited by the head. The machine halts if and when it reaches the special halt state 0. 

A universal Turing machine is capable of reading the transition rules of any other machine and performing the same computation over any initial configuration of the tape.

\subsubsection*{The halting problem}

One can ask whether there is a Turing machine $U$ which, given $code(T)$ and the input $s$, eventually stops and produces $1$ if $T(s)$ halts, and $0$ if $T(s)$ does not halt. Turing\cite{turing} proves that there is no such $U$.

\subsubsection*{The Busy Beaver game}

We denote by $(n,2)$ the class (or space) of all n-state 2-symbol Turing machines (with the halting state not included among the n states). A busy beaver machine\cite{rado} is a Turing machine that writes more 1s on the tape than any other of the same size (number of states).

If $\sigma_T$ is the number of 1s on the tape of a Turing machine $T$ upon halting, then:  $\sum(n)=\max{\{\sigma_T : T\in(n,2) \normalsize{\textbf{ }T(n)\textbf{ }halts}\}}$. 

If $t_T$ is the number of steps that a machine $T$ takes upon halting, then $S(n)=\max{\{t_T : T\in(n,2) \normalsize{\textbf{ }T(n)\textbf{ }halts}\}}$. $\sum(n)$ and $S(n)$ are noncomputable by reduction to the halting problem. Values are known for (n,2) with $n \leq 4$.

\subsection*{Algorithmic complexity}

The algorithmic complexity\cite{solomonoff,kolmo,levin,chaitin} (also known as Kolmogorov-Chaitin complexity or program-size complexity) $C_U(s)$ of a string $s$ with respect to a universal Turing machine $U$, measured in bits, is defined as the length in bits of the shortest Turing machine $U$ that produces the string $s$ and halts. Formally, $C_U(s) = \min\{|p|, U(p)=s\}$ where $|p|$ is the length of p measured in bits. 

Algorithmic complexity formalizes the concept of simplicity versus complexity. It opposes what is simple to what is complex or random.

\subsection*{Algorithmic probability}

Levin's $m(s)$ is the probability of producing a string $s$ with a random program $p$ when running on a universal prefix-free Turing machine\cite{levin}. That is, a machine for which a valid program is never the beginning of any other program, so that one can define a convergent probability the sum of which is at most 1. Formally, $m(s) = \Sigma_{p : U(p) = s} 2^{-|p|}$, i.e. the sum over all the programs for which $U$ with $p$ outputs the string $s$ and halts. $m(s)$ is the probability that the output of $U$ is $s$ when provided with a sequence of fair coin flip inputs as a program. $m$ is related to the concept of algorithmic complexity in that $m(s)$ is at least the maximum term in the summation of programs, which is $2^{-C(s)}$. Roughly speaking, algorithmic probability says that if there are many long descriptions of a certain string, then there is also a short description (low algorithmic complexity) and vice versa. As neither $C(s)$ nor $m(s)$ is computable, no program can exist which takes a string s as input and produces $m(s)$ as output.

\subsection*{String symmetry group}

The symmetry group of an object is the group of all isometries under which it is invariant. One can identify three symmetry preserving transformations for bit strings: identity (id), reversion (re), complementation (co) and the composition of (re) and (co) are the possible symmetry preserving operations. A tool that makes it possible to count the number of discrete combinatorial objects of a given type as a function of their symmetrical cases is provided by Burnside's lemma, given by the formula: $(2n + 2n/2 + 2n/2)/4$ for n odd, $(2n + 2(n+1)/2 )/4$ otherwise.


\newpage

\begin{thebibliography}{99}

\bibitem{bennett} Bennett, C.H. ``Logical Depth and Physical Complexity" in Rolf Herken (ed) \emph{The Universal Turing Machine--a Half-Century Survey,} Oxford University Press 227-257, 1988.
\bibitem{calude} Calude, C.S. \emph{Information and Randomness: An Algorithmic Perspective.} (Texts in Theoretical Computer Science. An EATCS Series), Springer; 2nd. edition, 2002.
\bibitem{caludemeyer} Calude, C.S. and Meyerstein, F. W. \emph{Is the universe lawful?} Chaos, Solitons \& Fractals 10, 6 (1999), 1075--1084.
\bibitem{chaitin} Chaitin, G.J. \emph{Algorithmic Information Theory.} Cambridge University Press, 1987.
\bibitem{delahaye} Delahaye, J.-P. and Zenil, H. ``On the Kolmogorov-Chaitin complexity for short sequences", in Calude, C.S. (ed.) \emph{Randomness and Complexity: from Chaitin to Leibniz.} World Scientific, p. 343--358, 2007.
\bibitem{kolmo} Kolmogorov, A.N. \emph{Three approaches to the quantitative definition of information}. Problems of Information and Transmission, 1(1): 1--7, 1965.
\bibitem{kirchherr} Kirchherr, W. Li, M. \emph{The miraculous universal distribution.} Mathematical Intelligencer, 1997.
\bibitem{levin} Levin, L. \emph{Laws of information conservation (non-growth) and aspects of the foundation of probability theory.} Problems in Form. Transmission 10, 206--210, 1974.
\bibitem{lloyd} Lloyd, S. \emph{Programming the universe.} Vintage, 2007.
\bibitem{rado} Rado, T. \emph{On noncomputable Functions.} \emph{Bell System Technical J.} 41, 877--884, May 1962.
\bibitem{pls} \emph{Pour La Science} (French edition of Scientific American), No. 400, February 2011.
\bibitem{shannon} Shannon, C. E. \emph{A Mathematical Theory of Communication.} The Bell System Technical J. 27, 379--423 and 623--656, July and Oct. 1948
\bibitem{solomonoff} Solomonoff, R. \emph{A Preliminary Report on a General Theory of Inductive Inference.} (Revision of Report V-131), Contract AF 49(639)-376, Report ZTB--138, Zator Co., Cambridge, Mass., Nov, 1960.
\bibitem{turing} Turing, A.M. \emph{On Computable Numbers, with an Application to the Entscheidungsproblem}, Proceedings of the London Mathematical Society. 2 42: 230--65, 1936, published in 1937.
\bibitem{wolfram} Wolfram, S. \emph{A New Kind of Science.} Wolfram Media, 2002.
\bibitem{zenil} Zenil, H. and Delahaye, J.-P. ``On the Algorithmic Nature of the World" in Dodig-Crnkovic, G. and Burgin, M. (eds.) \emph{Information and Computation.} World Scientific, 2010.

\end{thebibliography}
\end{document}